\documentclass[fleqn,10pt,twocolumn]{wlscirep}

\usepackage[table]{xcolor}
\usepackage[utf8]{inputenc}
\usepackage[T1]{fontenc}

\usepackage{amsmath}
\usepackage{amssymb}
\usepackage{glossaries}
\usepackage{graphicx}
\graphicspath{{Figures/}{./}}
\usepackage{here}
\usepackage{times}
\usepackage{units}
\usepackage{float}
\usepackage{booktabs}
\usepackage{bm}

\usepackage{multirow}
\usepackage{makecell}
\usepackage{tabularx}

\usepackage{mathtools}
\usepackage{xspace}
\usepackage{stfloats}
\usepackage{threeparttable}
\usepackage[none]{hyphenat}

\usepackage{caption}
\title{Dislocation-loop formation is a first-order phase transition}

\author[1]{Xiaoya Chang}
\author[2,3]{Arsalan Hashemi~\thanks{Corresponding author: arsalan.hashemi@uef.fi}}
\author[4,5]{Nima Ghafari Cherati}
\author[2,3]{Mikko Karttunen}
\author[4,5,6]{\'Ad\'am Gali~\thanks{Corresponding author: gali.adam@wigner.hun-ren.hu}}
\author[1,7]{Tapio Ala-Nissila~\thanks{Corresponding author: tapio.ala-nissila@aalto.fi}}

\affil[1]{MSP Group, Department of Applied Physics, Aalto University, P.O. Box 15600, FI-00076 Aalto, Espoo, Finland}
\affil[2]{European Laboratory for Learning and Intelligent Systems (ELLIS) Institute Finland, Maarintie 8, 02150 Espoo, Finland}
\affil[3]{Department of Technical Physics, University of Eastern Finland, P.O. Box 1627, FI-70211 Kuopio, Finland}
\affil[4]{HUN-REN Wigner Research Centre for Physics, P.O. Box 49, H-1525 Budapest, Hungary}
\affil[5]{Department of Atomic Physics, Institute of Physics, Budapest University of Technology and Economics, M\H{u}egyetem rakpart 3., H-1111 Budapest, Hungary}
\affil[6]{MTA--WFK Lend\"ulet ``Momentum'' Semiconductor Nanostructures Research Group, P.O. Box 49, H-1525 Budapest, Hungary}
\affil[7]{Interdisciplinary Centre for Mathematical Modelling and Department of Mathematical Sciences, Loughborough University, Loughborough, Leicestershire LE11 3TU, United Kingdom}

\begin{abstract}
Dislocation loops are the elementary product of radiation damage in crystals, limiting reactor-component lifetimes, power-electronics reliability and the coherence of solid-state qubits.
Their nucleation has been simulated for six decades but never reduced to a thermodynamic law.
We show that dislocation-loop formation is a \emph{first-order phase transition}, and construct its Ginzburg--Landau free energy, with the loop area as order parameter, entirely from atomistic simulation.
In diamond, carbon self-interstitials condense into planar precursors that collapse abruptly into a prismatic $\tfrac{1}{2}\langle110\rangle$ loop across a 3.7-electronvolt barrier, with pressure--volume work supplying only 2\% of the energy released.
The reduced free energy proves material-independent: the vacancy platelet-to-loop collapse in body-centred-cubic iron falls on the same one-parameter family, placing loop nucleation on a transferable thermodynamic footing.
\end{abstract}

\begin{document}

\flushbottom
\maketitle


Dislocation loops are the elementary unit of radiation damage in crystalline solids.
Condensing from supersaturated point defects, they nucleate, grow and interact to produce the swelling and embrittlement that limit the lifetime of fission- and fusion-reactor components, they degrade high-power semiconductor devices, and they decohere solid-state quantum bits~\cite{arakawa2007observation,yi2013insitu,shikata2022diamond,yan2024multi}.
Six decades of transmission electron microscopy and atomistic simulation have established what loops look like, how they migrate and what they cost energetically~\cite{osetsky2000stability,marian2002mechanism,peng2018shockwave,ma2020multiscale,gao2021mechanisms}.
How they are \emph{born}, however, is still described kinetically rather than thermodynamically: loop nucleation is treated as a rate process, not as a phase transition possessing an order parameter, a free energy and a barrier.
Two long-standing controversies show the cost of this gap.
The origin of the $\{001\}$ platelets of type-Ia diamond has been debated since the 1960s~\cite{goss2003extended,Humble_physsci_1982,fallon1995nitrogen,sumida1988measurement,baker1998new}, and the near-equal populations of $\langle100\rangle$ and $\tfrac12\langle111\rangle$ loops in ferritic steels have resisted explanation for almost as long~\cite{marian2002mechanism,yang2024formation}; in both cases the state of the art remains a schematic reaction-energy landscape.

Diamond is a stringent test bed for this question: it is interstitial-dominated, free of the magnetic complications of $\alpha$-iron, and its extended defects have been characterised for sixty years.
Even the most advanced growth techniques leave intrinsic defects behind~\cite{nemeth2020complex}, and while individual interstitials can support optically active quantum states, their aggregation produces extended defects that degrade diamond's optical and electronic performance~\cite{shikata2022diamond,yan2024multi}.
At elevated interstitial densities, where defect mobility may be suppressed, diamond develops the extended planar defects known as platelets, whose formation has long been attributed to extrinsic species such as nitrogen~\cite{goss2003extended,olivier2018imaging}.
Although the core structures, energies and glide of individual dislocations in diamond are well established from electronic-structure calculations~\cite{blumenau2002dislocations,blumenau2003dislocations,chen2025electronic}, the atomistic mechanism by which interstitials aggregate into extended defects has remained unresolved.

Self-interstitials in diamond have received comparatively little attention, largely because they are widely assumed to become mobile above 700\,K and to recombine upon thermal treatment~\cite{Hunt2000PRB,Kiflawi2007JPCM}.
Under interstitial-rich conditions, however, they may instead aggregate, self-organizing into line motifs and ultimately into dislocation loops and platelet-like structures.
Testing whether such extended defects can form \emph{intrinsically} -- without nitrogen -- requires nanometre-scale supercells and nanosecond trajectories at near-quantum-mechanical accuracy, which we reach using a machine-learned interatomic potential benchmarked against density functional theory in earlier work~\cite{chang2026machine,song2024general,xu2025gpumd} (Supplementary Materials).

We find that carbon self-interstitials aggregate by diffusion--recombination and lattice exchange into recurrent line-defect motifs, and that these reorganize into a prismatic $\tfrac{1}{2}\langle110\rangle$ dislocation loop and two platelet-like planar defects -- all in nitrogen-free diamond.
The final step is abrupt: the potential energy and the pressure drop discontinuously, the system selects a single $\{110\}$ habit plane, and the transition admits a Ginzburg--Landau (GL) description with the loop area as its order parameter.
Fixing every coefficient from the simulations shows the transition to be strongly first-order and driven overwhelmingly by local bond reorganization, across a barrier of $3.7$~eV whose Arrhenius extrapolation to 2800~K coincides with the range over which diamond platelets are observed to convert into interstitial loops~\cite{evans1995conversion}.
Because the resulting reduced free energy retains a single free parameter, it is not specific to diamond: the vacancy platelet-to-loop collapse in body-centred-cubic iron~\cite{yang2024formation} falls on the same one-parameter family. 

\section{Results}
\label{sec:results}

\subsection{Aggregation upon annealing}
\label{subsec:aggre_mechasim}

\begin{figure*}[b!]
    \centering
    \includegraphics[scale=0.53]{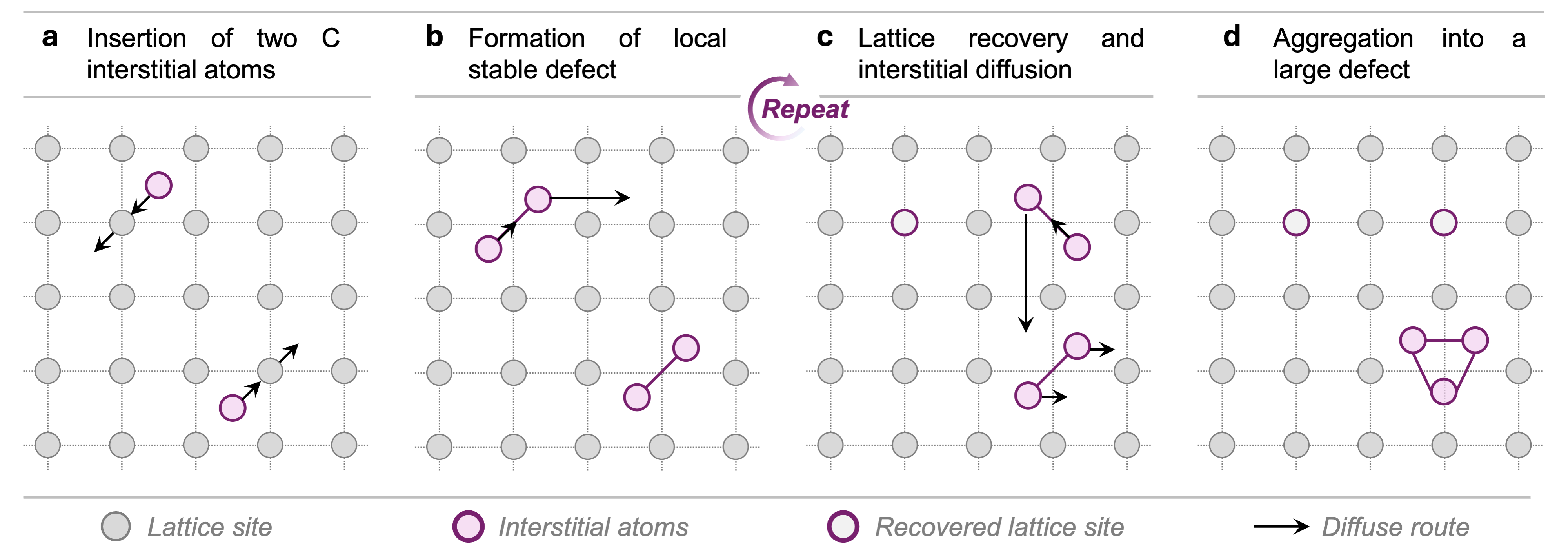}
    \captionsetup{justification=justified}
    \caption{\textbf{Schematic of diffusion and lattice-exchange mechanism for carbon interstitial aggregation in diamond.}
    A connection between two purple atoms indicates a local stable split defect, whereas three connected atoms correspond to an aggregated cluster.}
    \label{fig:aggregation}
\end{figure*}
%
While our previous work~\cite{chang2026machine} focused on identifying new defect configurations (C$_\text{2}$ to C$_\text{8}$) through annealing MD simulations, here we investigate how initially dispersed carbon interstitials aggregate into stable and metastable defect configurations.
The annealing pathway leading to the recurrent line-defect motif C$_\text{4A}$ (whose isolated structure is shown in Fig.~S1 and fully characterized in Ref.~\citeonline{chang2026machine}) is resolved at the atomic level in Fig.~S2, where atoms displaced from ideal lattice positions are highlighted for clarity.
Throughout this evolution, interstitial atoms undergo thermally activated migration, with half becoming trapped at lattice sites and the others diffusing over long distances, accompanied by an abrupt increase in atomic displacement (Fig.~S2b).
Importantly, the resulting defect is not formed solely from the initially introduced self-interstitials; instead, most of its constituent atoms originate from lattice sites (Fig.~S2c,d), indicating that defect formation proceeds through lattice atom participation and reconstruction.

The formation of other reported interstitial defects~\cite{chang2026machine} follows the same aggregation mechanism.
Figure~\ref{fig:aggregation} summarizes the thermally activated evolution of dispersed carbon interstitials into stable defect configurations.
At the early stage, interstitial atoms preferentially interact with nearby lattice sites, forming energetically favorable local defect motifs such as split-interstitial pairs.
With increasing temperature, thermal activation enables one atom in a split-interstitial pair to escape and become mobile, while its partner remains near the lattice site.
The mobile interstitial then diffuses through the lattice and may recombine with another lattice site, generating a locally stable defect similar to the initial configuration.
This diffusion-recombination cycle can repeat multiple times during annealing.
Once a mobile interstitial is incorporated into an existing defect complex, it becomes immobilized, and further evolution proceeds primarily through local rearrangements within the defect rather than long-range diffusion.
The above findings support the third mechanism discussed by Goss et al.~\cite{goss2003extended}, involving a concerted exchange between the impurity and neighboring lattice atoms, without requiring a vacancy or an impurity interstitial.

By tracking the atomic displacements before and after annealing (see Fig.~S3), the extent of atomic participation in these processes can be quantified.
As the number of interstitial atoms increases, progressively larger fractions of the surrounding lattice become involved in defect formation and rearrangement (Fig.~S4), indicating that defect evolution becomes increasingly collective. 
For dense interstitial clusters such as C$_8$, an average of 10.6\% of lattice atoms participates in the rearrangement, implying increasingly cooperative atomic motion and a progressively more demanding kinetic landscape, likely involving higher activation barriers and longer annealing times.
To access these rare activated processes within accessible MD timescales, annealing simulations were performed at elevated temperatures (\emph{e.g.}, 4500\,K) as an accelerated-sampling strategy.
The simulated temperatures should therefore be interpreted as sampling conditions rather than experimental annealing temperatures.
Under experimentally relevant conditions, limited interstitial mobility may kinetically trap isolated defects or small aggregates, whereas elevated temperatures, pressure, strain, or pre-existing defects may promote their evolution into larger complexes, including planar defects~\cite{goss2001interstitial}.

With increasing interstitial content, the aggregated clusters evolve into a recurrent zigzag chain-like motif that elongates while preserving its topology (Fig.~S5).
Projected onto the $(011)$ plane, this family of line defects comprises a chain of six-membered carbon rings flanked by alternating five- and seven-membered rings.
This motif, derived from the C$_\text{4A}$ building block~\cite{chang2026machine} (Fig.~S1), provides the structural template from which the extended defects analyzed next emerge.

\subsection{Extended defects}
\label{subsec:extended_defect}

\begin{figure*}[t!]
    \centering
    \includegraphics[scale=1.05]{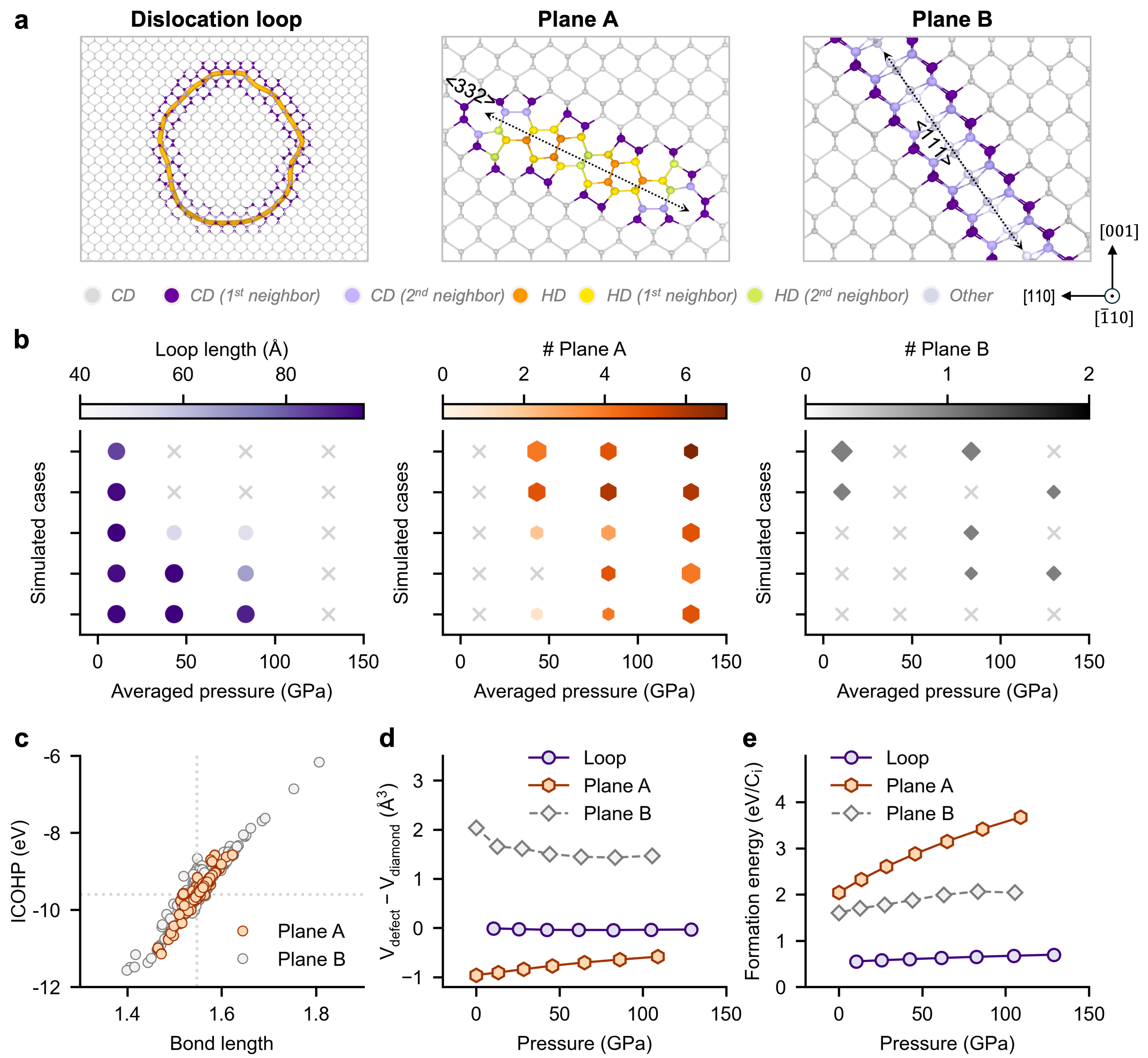}
    \captionsetup{justification=justified}
    \vspace{-0.3em}
    \caption{
    \textbf{Visualisation of extended defects and their pressure dependency.}
    (a) Generated dislocation loop, plane~A, and plane~B, coloured by their local structural environments. CD and HD denote cubic diamond and hexagonal diamond, respectively, as identified by OVITO\cite{stukowski2010visualization}.
    (b) Formation frequency of dislocation loops, plane~A, and plane~B under different pressures across five independent simulations.
    The size and color intensity of the purple markers correspond to the dislocation loop density.
    In contrast, for the orange hexagonal and grey rhombi markers, the color intensity represents the number of planar defects, while the marker size indicates their average size.
    Grey cross markers represent the absence of the corresponding defects.
    (c) Calculated ICOHP values for each C--C bond within the two planar defects as a function of bond length. For reference, the dot lines indicate the ICOHP value (-9.59~eV) and bond length (1.547~\AA) of pristine diamond.
    (d-e) Average volume difference (between defective and bulk diamond atoms) and formation energy of the three defect types induced by a single interstitial atom.}
    \label{fig:three_defect}
\end{figure*}

Carbon self-interstitials aggregate into platelet-like extended defects in nitrogen-free diamond, bearing directly on the platelet-origin debate outlined above.
To capture extended defect formation, we enlarged the simulation cell to 5~nm, containing up to 27,000 carbon atoms and 300 interstitials, corresponding to an interstitial density of 2.4~nm$^{-3}$.
High-temperature simulations revealed three classes of extended defects: a predominant dislocation loop and two planar structures, here termed plane~A and plane~B (Fig.~\ref{fig:three_defect}a).
In several simulations, the resulting defect complexes comprised multiple extended defects (Fig.~\ref{fig:three_defect}b), indicating that defect evolution proceeds through cooperative aggregation rather than the formation of isolated defects.
All three structures retain predominantly \(sp^3\)-bonded character; extraction details and atomic coordinates are given in the SI.

The dislocation loop lies approximately in the $\{110\}$ plane and closely resembles a prismatic loop.
Consistent with previous reports~\cite{pirouz1983dissociation,nie2020direct,li2023dislocation}, the loop is characterized by a Burgers vector of $\frac{1}{2}a_0\langle 110\rangle$.
Transmission-electron-microscopy studies of natural type-IaB diamond establish that the dislocation loops formed there are prismatic and, in every case tested, of \emph{interstitial} character~\cite{hirsch1986platelets,evans1962dislocation} -- matching the interstitial nature of the loop formed here.
Those loops carry Burgers vectors $a_0\langle001\rangle$ (majority) or $\tfrac{1}{2}a_0\langle110\rangle$, and arise from the high-temperature transformation of nitrogen platelets into interstitial loops and voidites~\cite{hirsch1986platelets,evans1995conversion}.
Our nitrogen-free simulations reproduce the $\tfrac{1}{2}a_0\langle110\rangle$ interstitial-loop family and the underlying planar-defect$\to$loop transformation \emph{intrinsically} -- without the nitrogen platelets that drive it experimentally -- consistent with the complementary intrinsic pathway proposed above.
The experimentally resolved loops are, however, tens to hundreds of nanometres across, so the nanometre-scale loop obtained here (radius $\approx 1.5$~nm, set by the accessible supercell) represents the early, nucleation-stage end of this loop family.
Because a prismatic interstitial loop is a condensed disc of self-interstitials, its area counts the interstitials it has captured, $N_i=\pi R^2 b/\Omega$ -- a relation our simulations verify directly, predicting $317$ interstitials for the loop obtained here against the $300$ actually introduced.
Applying it to the observed loop population of Hirsch \emph{et al.}~\cite{hirsch1986platelets} (diameters $\approx10^2$~nm at a concentration $\approx10^{19}$~m$^{-3}$) implies an interstitial content of $\approx30$--$700$~ppm in such diamonds -- the same order as their aggregated-nitrogen budget and consistent with a platelet-dissociation origin (see SI for details).
Nearly fifty years ago, Woods~\cite{woods1976electron} suggested that dislocation loops form through interactions among adjacent platelet defects, with the resulting loops helping to relax the surrounding strain field.
This proposed mechanism aligns with our simulations, where we find that dislocation-loop formation is energetically favourable and effectively alleviates internal stress (Fig.~\ref{fig:loop_formation}).

Plane A consists of a hexagonal backbone extending along $\langle 332\rangle$.
Viewed edge-on, it reduces to a six-membered interstitial chain symmetrically flanked by five- and seven-membered rings, consistent with the C$_{\mathrm{4A}}$-containing motifs reported in silicon~\cite{lee2009integrated}.
This suggests that the C$_{\mathrm{4A}}$-derived line defect (Fig.~S5a) serves as the fundamental building unit of plane~A.
Plane~B closely resembles the Humble platelet model for nitrogen-containing diamond~\cite{humble1985platelet}: a sheet of puckered pentagonal C$_{\mathrm{6A}}$ units (Fig.~S1) parallel to the $\{100\}$ planes.
Adjacent C$_{\mathrm{6A}}$ units generate locally octahedral carbon coordination (Fig.~S7), consistent with the structure reported by Botti \emph{et al.}~\cite{botti2013carbon}.
Particularly, electronic structure analysis reveals that plane~A does not introduce any states within the band gap, whereas plane~B, containing dangling bonds, exhibits localized defect states and may become optically active (Fig.~S8).
Analogous localized -- and even one-dimensional -- electronic states have been predicted at dislocation cores in diamond~\cite{polatgenlik2022dislocations}, of potential relevance to proximate quantum defects.
Nitrogen is therefore not a prerequisite for these planar defects, although in natural type-Ia diamond it may still template their formation.

We next examine the pressure dependence across five independent simulations, identifying each defect type individually since several can coexist within a single complex (Fig.~\ref{fig:three_defect}b).
The dislocation loop is favoured at relatively low pressures, whereas the hexagonal plane~A is more easily generated under extreme compression~\cite{wang2023dislocation}.
Additionally, plane~B appears in limited simulations, which suggests it is more likely to be a secondary feature rather than a dominant product.
Their contrasting pressure responses are rooted in distinct bonding characteristics.
DFT-based ICOHP analysis (Fig.~\ref{fig:three_defect}c) shows that plane~A has a narrow distribution of both bond lengths and bond strengths about the pristine-diamond values, whereas plane~B is substantially broader, indicating greater structural distortion and bonding heterogeneity (SI).
Plane~A is accordingly more robust, consistent with its higher prevalence under elevated pressure.
Although both planar defects emerge naturally through self-interstitial aggregation in our simulations, whether either represents the atomic structure of the platelet defects observed experimentally remains an open question.

Only these three classes of extended defect were observed under the interstitial-rich conditions considered here; others may require extrinsic species or different growth environments.
They differ markedly in the extent of lattice reconstruction, with progressively more of the surrounding lattice involved on passing from the loop to the planar defects (Fig.~S9).
Their distinct pressure responses follow from the atomic volumes and formation energies under compression (Fig.~\ref{fig:three_defect}d,e; SI): the loop retains an atomic volume close to that of pristine diamond together with the lowest and most weakly pressure-dependent formation energy, whereas both planar defects are penalized under compression.
The competition is therefore between formation energy and excess volume -- low-energy configurations such as the loop are favoured near ambient conditions, while compact, low-excess-volume structures such as plane~A become increasingly favourable under pressure.

\subsection{Loop formation as a first-order phase transition}
\label{subsec:GL}

\begin{figure*}[h]
    \centering
    \includegraphics[width=1\linewidth]{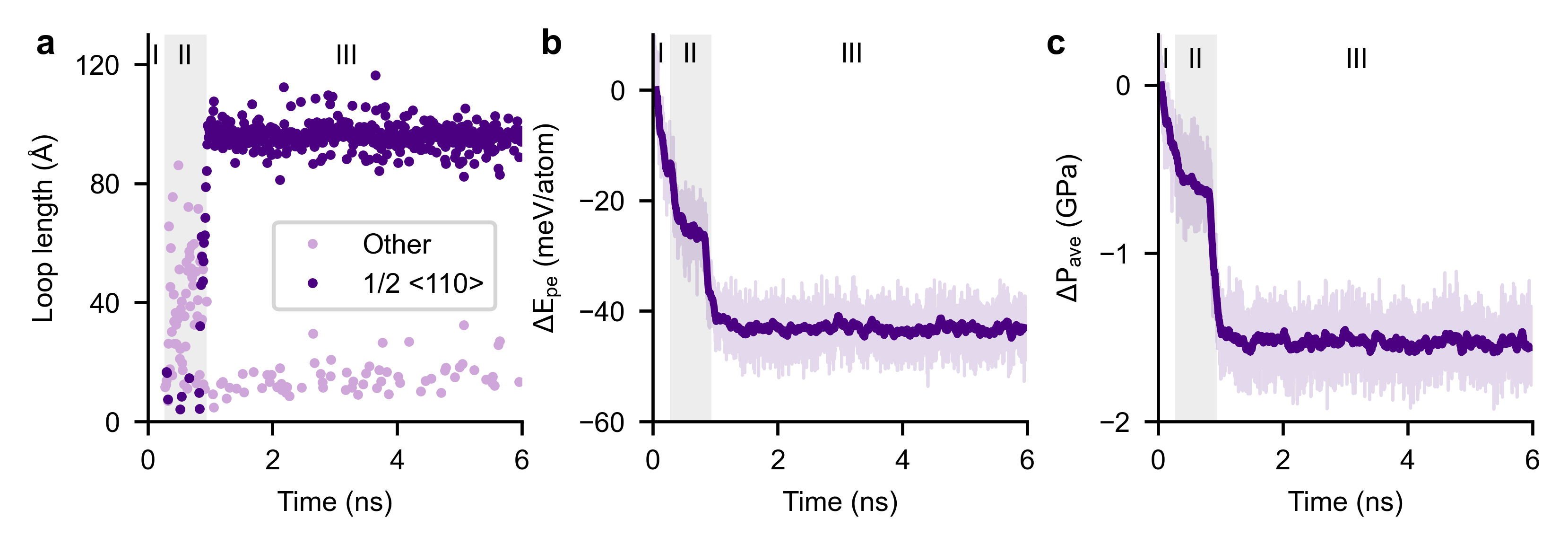}
    \captionsetup{justification=justified}
    \vspace{-0.7cm}
    \caption{\textbf{Dislocation-loop formation is a first-order transition.}
    Characterization of the three-stage loop-formation process during heating at 6000~K.
    (a) Dislocation-loop types and their lengths, with both unclassified loops (``other'') and $\tfrac{1}{2}\langle110\rangle$ loops identified by dislocation extraction analysis (DXA) in OVITO~\cite{stukowski2010visualization}.
    (b) Potential-energy change ($\Delta E_{\text{pe}}$) and (c) average pressure ($\Delta P_{\text{ave}}$) during the stage II and III; both drop abruptly once the $\tfrac{1}{2}\langle110\rangle$ loop forms after Stage~II, the signature of a first-order transition.
    The three formation stages are indicated and the light-grey region highlights Stage~II.}
    \label{fig:loop_formation}
\end{figure*}

The large-scale simulations reveal that the dislocation loop emerges through three stages: interstitials first aggregate into numerous small planar defects (Stage~I); these coalesce into unclassified, highly distorted loops (Stage~II, the line-defect phase); and finally, the system abruptly transforms into a stable prismatic $\tfrac{1}{2}\langle110\rangle$ loop (Stage~III, the loop phase) (Fig.~\ref{fig:loop_formation}).
At the Stage~II$\to$III transition, the potential energy and average pressure drop simultaneously and discontinuously (Fig.~\ref{fig:loop_formation}), marking the hallmark of a \emph{first-order phase transition}.
The transition also breaks symmetry: from a line-defect phase with no preferred plane, the system selects a specific $\{110\}$ plane and Burgers-vector direction.

These features motivate a Ginzburg--Landau (GL) description with the loop area $\psi=A$ as the order parameter (zero in the line-defect phase, $\psi_0$ in the loop phase). The free energy
\begin{equation}
  \mathcal{F}(\psi) = \tfrac{a}{2}\psi^2 + \tfrac{b}{4}\psi^4 + \tfrac{c}{6}\psi^6 - f_{\mathrm{ext}}\psi,
  \label{eq:GL_main}
\end{equation}
with $a>0$, $b<0$, $c>0$, is the canonical sextic construction for a first-order transition, where $f_{\mathrm{ext}}$ is the elastic-stress coupling (Supplementary Materials).
Parameterizing Eq.~\eqref{eq:GL_main} using the averaged results from three independent simulations (loop area $\psi_0\approx702$~\AA$^2$, from an extracted loop radius of $14.95$~\AA; free-energy well depth $\Delta F\approx678.7$~eV; pressure--volume energy $\Delta F_{PV}\approx13.7$~eV) yields three central results.
First, the pressure--volume work released on loop formation accounts for only $\tilde f = \Delta F_{PV}/\Delta F \approx 2\%$ of the total energy release: the remaining $\approx98\%$ originates from local bond reorganisation rather than from long-range elastic relaxation.
Second, the transition is strongly first-order by its order-parameter discontinuity ($\psi:0\to702$~\AA$^2$), yet the nucleation barrier is small: matching the observed nanosecond-scale transition at 6000~K to Arrhenius kinetics gives $\Delta F^{*}\approx3.7$~eV ($\ll\Delta F$), corresponding to a critical loop nucleus of $\approx150$~\AA$^2$ ($R^{*}\approx7$~\AA).
The line-defect phase is therefore only weakly metastable, sitting just above the spinodal, which explains why the transformation appears so abrupt. 
Dislocation nucleation in diamond has likewise been examined by atomistic and DFT-MD methods~\cite{yang2018homogeneous,chen2025electronic}, consistent with a kinetically controlled barrier.

We can further use the MD parameters to estimate the relevant time scales from Kramers' rate theory~\cite{Kramers1990} in the overdamped limit. 
We find that the relaxation time in the metastable basin is $\tau_{\mathrm{rel}}\approx0.20$\,ps, comparable to characteristic phonon timescales in diamond.
Thus, local equilibration of the order parameter is rapid, and the observed nanosecond-scale loop-formation times are dominated by the activated crossing of the free-energy barrier.
A full derivation, the parameterization and the kinetic determination of the barrier are given in the Supplementary Materials.

\begin{figure*}[t!]
    \centering
    \includegraphics[width=1\linewidth]{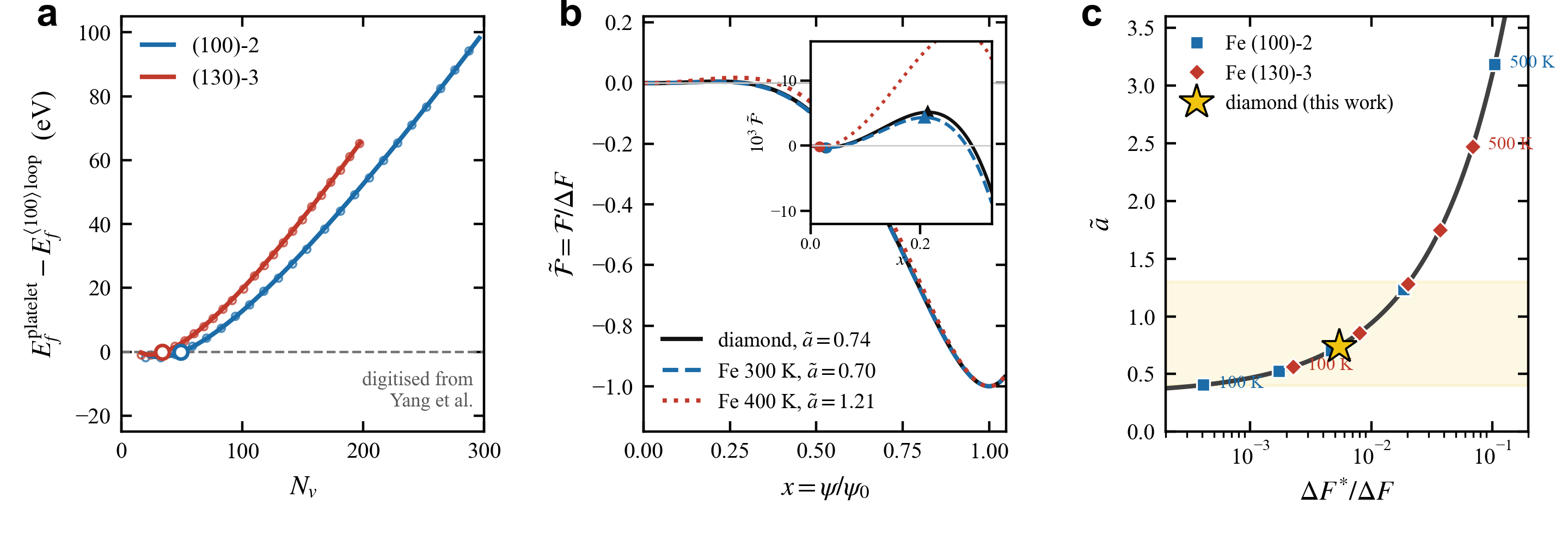}
    \captionsetup{justification=justified}
    \caption{\textbf{The reduced free energy is material-independent: diamond and bcc iron on the same one-parameter family.}
    (a) In bcc iron, the difference in formation energy between a two-dimensional vacancy platelet and a $\langle100\rangle$ prismatic vacancy loop against vacancy number $N_v$, for platelets of two and three atomic layers on $(100)$ and $(130)$ planes (open circles, digitised from Yang \emph{et al.}~\cite{yang2024formation}; lines, fits). This difference is the iron analogue of $\Delta F$. Open markers on the dashed line mark the thermodynamic crossover above which the loop is favoured, reproducing the reported critical sizes of $47$ and $33$ vacancies to within $3$--$6\%$.
    (b) Reduced free energy $\tilde{\mathcal F}(x)=\mathcal{F}/\Delta F$ against the normalised loop area $x=\psi/\psi_0$ for diamond ($\tilde a=0.74$, this work) and for iron at $300$ and $400$~K; the diamond and $300$~K iron curves are nearly coincident. Inset: the barrier region magnified, showing the metastable planar-precursor minimum (circles) and the critical nucleus (triangles).
    (c) The one-parameter mapping $\tilde a(\Delta F^{*}/\Delta F)$ that follows from the two data constraints, with the iron loci over $100$--$500$~K (squares, diamonds) and the diamond value (star). Both materials fall in the band $\tilde a=\mathcal{O}(1)$ (shaded), despite opposite defect character, different lattices and bonding, and barriers and well depths differing by more than an order of magnitude.}
    \label{fig:GL_universal}
\end{figure*}

Two features make this description general rather than diamond-specific.
First, because the elastic coupling is weak, the two data constraints collapse onto a material-independent one-parameter family, $\tilde b=-(12+4\tilde a)$ and $\tilde c=12+3\tilde a$, for which the metastability discriminant becomes an identity, $\tilde b^2-4\tilde a\tilde c=4(\tilde a+6)^2>0$.
The first-order character is therefore \emph{structural}: given a stationary loop state and a finite energy release, a metastable minimum, a barrier and an order-parameter discontinuity follow necessarily, and the single remaining parameter is fixed by the dimensionless ratio $\Delta F^{*}/\Delta F$ alone.
Second, this reduction carries over to a very different material.
The collapse of two-dimensional vacancy platelets into $\langle100\rangle$ prismatic loops in bcc iron~\cite{yang2024formation} has the same topology as the transition analysed here -- a planar precursor converting abruptly into a loop above a critical size -- and is itself a long-debated problem in the principal structural material for nuclear systems~\cite{marian2002mechanism}.
Combining the published formation energies of the two competing iron phases with the reported critical sizes places iron at $\tilde a=0.4$--$3.1$ between 100 and 500~K, straddling the diamond value of $0.74$, with $\tilde a=0.70$ for the $(100)$-2 platelet at 300~K (Fig.~\ref{fig:GL_universal}; SI, Sec.~\emph{Transferability test}).
The two systems thus occupy the same narrow stretch of the universal family despite opposite defect character, different lattices and bonding, and barriers and well depths that differ by more than an order of magnitude.
These estimates are order-of-magnitude, since the iron barrier is inferred from a threshold criterion rather than measured directly; nevertheless they indicate that the same reduced description should apply to the irradiation-induced loops that govern microstructural evolution in nuclear materials~\cite{arakawa2007observation,yi2013insitu}.

Landau and phase-field descriptions of dislocations and defect microstructure are, of course, well established -- including phase-field models of microstructure evolution~\cite{chen2002phasefield}, phase-field dislocation dynamics~\cite{wang2001nanoscale}, the phase-field-crystal model of dislocation nucleation~\cite{elder2004modeling}, and Landau theories of crystal plasticity~\cite{salman2011minimal}.
These are, however, typically spatially-resolved field theories evolved dynamically with phenomenological or elasticity-derived coefficients.
Here, by contrast, we construct a minimal Landau free energy whose single geometric order parameter (the loop area) and coefficients are fixed directly by atomistic simulation, reducing an explicitly resolved atomistic transition to a closed thermodynamic description and characterizing loop formation as a first-order transition.

\section{Discussion}
\label{sec:discussion}

The first of the two controversies set out above can now be reframed.
Plane~A reproduces the silicon-platelet structural family, whereas plane~B matches the Humble model previously associated with nitrogen-containing diamond -- yet both arise here without nitrogen, so carbon-interstitial aggregation alone is a sufficient route to platelet-like planar defects, complementary to the nitrogen-mediated mechanisms proposed for natural type-Ia diamond.
This reframes rather than closes the decades-old platelet-origin debate.
Consistent with this, the interstitial loops observed in nitrogen-bearing diamond are predominantly $a_0\langle001\rangle$ and form by platelet degradation~\cite{hirsch1986platelets,evans1995conversion}, whereas the carbon-only route modelled here yields the $\tfrac{1}{2}a_0\langle110\rangle$ family; the difference in loop character plausibly reflects the distinct formation pathways.
Which extended defect prevails, we find, is set not only by formation energies but also by volumetric relaxation, which governs their pressure-dependent competition.

The reduction itself applies to any defect-morphology change that proceeds discontinuously and admits a geometric order parameter: given measurable drops in energy and stress, such a transition can be cast as a minimal Landau free energy parameterized from simulation rather than from phenomenology.
That the bcc-iron platelet-to-loop collapse falls on the same one-parameter family, despite opposite defect character and a barrier twenty-five times smaller, shows that the description is tied to neither a single material nor a single type of bonding.
The nanometre loop obtained here is moreover directly comparable in size to the interstitial loops characterised in irradiated metals -- from the $\tfrac{1}{2}\langle111\rangle$ loops imaged in $\alpha$-Fe~\cite{arakawa2007observation} to those in self-ion-irradiated tungsten~\cite{yi2013insitu} -- whose energetics are well established from atomistic simulation with empirical many-body potentials, and which are here recast as a first-order transition.
The framework thereby converts large-scale atomistic trajectories into closed-form thermodynamic descriptions of microstructural evolution, spanning dislocation loops, platelets, stacking faults and precipitate transformations alike.

\section{Materials and Methods}
Full details of the density functional theory calculations, the large-scale machine-learned molecular dynamics protocol, and the Ginzburg--Landau and Kramers' parameterization are given in the Supplementary Materials.

\section{Acknowledgments}
\label{acknowledgments}

This work was supported by the Academy of Finland through grants 370057 and 373647, and by the European Union and the European Innovation Council through the Horizon Europe project QRC-4-ESP (Grant No. 101129663), and EU Horizon Europe Quest project (No. 10116088).
A.H. and M.K. were supported by the Research Council of Finland (Flagship of Advanced Mathematics for Sensing Imaging and Modeling, grant 358944) and by Foundation PS.
We gratefully acknowledge CSC--IT Center for Science Ltd., Finland, and the Aalto Science--IT project for providing the computational resources and computing time used in this work.
A.G. acknowledges the high performance computational resources provided by KIF$\text{\"{U}}$ (Governmental Agency for IT Development of Hungary) and the European Commission for the projects SPINUS (Grant No.\ 101135699) and QuSPARC (Grant No.\ 101186889).

\section{Author contributions}
\label{contribution}

\textbf{X. C.} performed the large-scale simulations, formal analysis, visualisation, and wrote the original manuscript draft.
\textbf{A. H.} contributed to conceptualization, methodology, validation, supervision, and extensively reviewed and edited the manuscript.
\textbf{N. G. C.} examined the generated structures and reviewed the manuscript.
\textbf{M. K.} contributed to technical aspects, and reviewed and edited the manuscript, and acquired funding.
\textbf{A. G.} developed the Ginzburg--Landau analysis, reviewed and edited the manuscript, provided suggestions and acquired funding.
\textbf{T. A.-N.} developed the Kramers' analysis, supervised the project, acquired funding, provided resources, managed project administration, and reviewed and edited the manuscript.

\section{Data availability}
\label{data}

All data supporting the findings of this study are available within the paper and its Supplementary Information. 
Simulation input files, atomic coordinates of the newly identified defect structures, and the code used for the GL–Kramers analysis will be made publicly available through a GitHub repository upon acceptance of the manuscript.

\section{Competing interests}
The authors declare no competing interests.

\bibliography{main.bib}
\end{document}